\documentstyle[aps,manuscript,epsfig]{revtex}

\newcommand{\etal}{\textit{et al.}}
\newcommand{\nima}{Nucl. Instr. Meth. A}

\newcommand{\prcc}{Phys. Rev. C}
\newcommand{\plb}{Phys. Lett. B}
\newcommand{\npa}{Nucl. Phys. A}
\newcommand{\epja}{Eur. Phys. Jour. A}
\newcommand{\zpa}{Z. Phys. A}

\begin{document}

\draft

\title{Experimental studies of unbound neutron-rich nuclei}

\author{J.~L.~Lecouey\thanks{\textit{E-mail address:} 
 lecouey@nscl.msu.edu \newline \textit{Present address:} National Superconducting 
 Cyclotron Laboratory, Michigan State University,  
 East Lansing, MI 48824, USA } \\ for the LPC-CHARISSA-DEMON Collaboration}

\address{Laboratoire de Physique Corpusculaire de Caen, IN2P3-CNRS, \\
ENSICAEN et Universit\'e de Caen, 6 boulevard Mar\'echal Juin, \\
14050 Caen cedex, France}

\maketitle

\begin{abstract}
     The three-body description of two-neutron halo nuclei relies on the two-body 
     interactions between the constituents. In order to provide constraints on 
     calculations devoted to $^{14}$Be and $^{17}$B, the neutron unbound states of 
     $^{13}$Be and $^{16}$B have been investigated by one-proton knockout. The 
     experimental techniques and results are discussed here.

\end{abstract}

\bigskip

\section{Introduction}

In recent years nuclei at the limit of stability have become 
experimentally accessible. In these regions where systems present an 
excess of neutrons or protons, the traditional mean-field picture is often no longer valid and 
the nucleus can be more appropriately described as a few-body system. 
Halo nuclei, in which one or two nucleons extend far from the rest of the nucleus (the core), are a typical example of such a structure.

On the neutron-rich side there exist three-body (core-neutron-neutron) systems (Borromean) which are probably the most intriguing halo nuclei: they are bound while none of 
the two-body subsystems ($n$-$n$ and core-$n$) are \cite{Han95}.  In these cases, 
the structure of the latter is known to play a significant role in the 
description of the halo nucleus. Thus $^{11}$Li cannot be described without 
assuming the presence of a virtual $s$-state in $^{10}$Li \cite{Tho94}. 
In the same way, $^{13}$Be has been investigated due to the importance of a 
possible $s$-state near threshold for the structure of the two-neutron halo
nucleus $^{14}$Be \cite{Tho96}.

Since the measurement of its interaction cross-section \cite{Tan90}, $^{17}$B has 
been considered to be the heaviest two-neutron halo
nucleus. Little is known about the unbound $^{16}$B other thant that the (probable) ground
state was weakly populated in a multinucleon transfer experiment at HMI \cite{Kal00}.
In order to improve our knowledge of the spectroscopy of $^{16}$B and $^{13}$Be the experiments described here were carried out.

\section{Experimental techniques}

Historically neutron unbound states (resonances) close to the stability line were studied by using beams of neutrons \cite{Hae83}. Obviously this technique is no longer practical near the dripline given that the target itself is $\beta$-unstable
(the half-lives of $^{12}$Be and $^{15}$B are 24 and 10.4 ms).

Multinucleon transfer reactions have been used extensively to access exotic unbound systems. 
However the reaction mechanism is complex and the missing mass spectrum often exhibits a large background.
Unbound states of exotic nuclei can also be populated by removal (or ``knockout'') of one or several nucleons from a high-energy beam.
The reconstructed fragment-neutron relative energy (or velocity) spectrum will exhibit structures which correspond to the energy of the
populated unbound states with respect to the neutron decay threshold.

More recently, two-proton knockout has been used to study $^9$He \cite{Che01}. The 
use of a simpler reaction than projectile fragmentation (e.g. \cite{Thn00})
is advantageous as the backgrounds are suppressed and as
to first order the neutron configuration remains undisturbed by the 
reaction, the final-state neutron angular momentum is known,
provided the structure of the projectile is known.

For $^{13}$Be and $^{16}$B one-proton knockout reactions were chosen. 
The experiments were performed at GANIL, using a 35 (41) MeV/nucleon $^{17}$C
($^{14}$B) beam impinging on a C target. 
The charged fragments arising from the reactions were detected using a 
Si-Si-CsI telescope located some 15~cm  downstream of the target.
The two silicon detectors provided for good position information (1 mm), and combined 
with the energy signals derived from the CsI, unambiguous identification of the fragment.
The total energy resolution of the telescope was 1.2\% ({\sc fwhm}).

The neutrons were detected using 97 elements of the {\sc demon} array \cite{Til95}. 
The liquid scintillator modules were arranged
in a staggered configuration covering polar angles up to 39$^\circ$ in the 
laboratory, providing for a 
significant detection efficiency up to a few MeV relative energy (figure \ref{fig:simul}). 
The neutron energy was deduced from the time-of-flight with a resolution of 5\%.

\section{Results}

\begin{figure}
\begin{center}
\mbox{\psfig{file=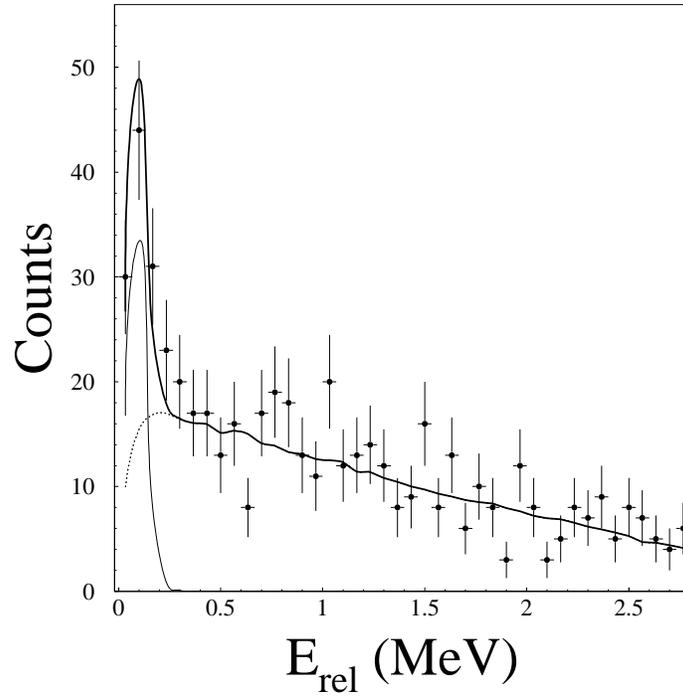,width=9.5cm}}
\caption{$^{15}$B-$n$ relative energy spectrum. The points are the data, the thick solid line the result of a fit including a $d$-wave resonance
(thin solid line) and an event-mixing distribution (dotted line).}
\label{fig:16B}
\end{center}
\end{figure}

\begin{figure}
\begin{center}
\mbox{\psfig{file=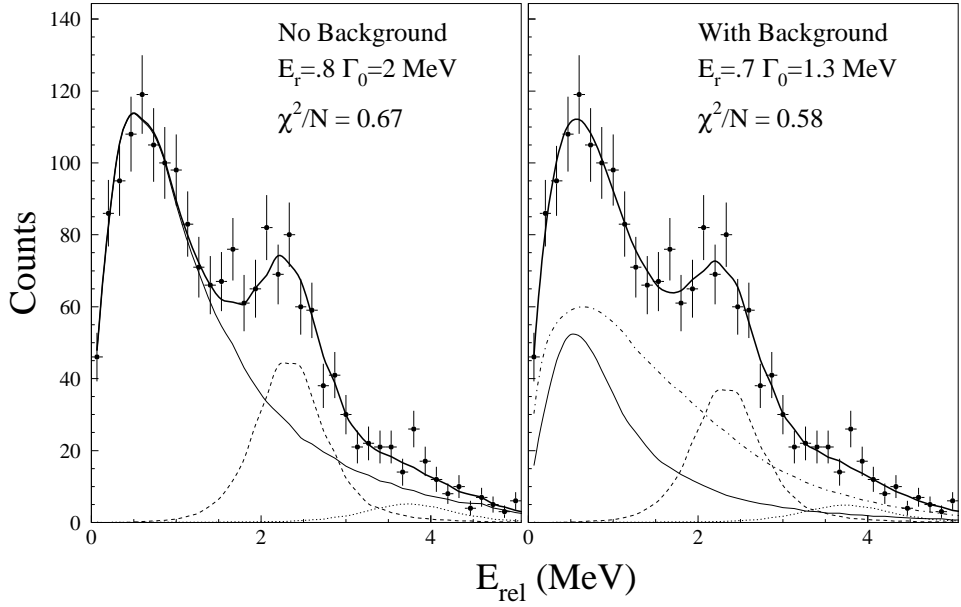,width=14cm}}
\caption{$^{12}$Be-$n$ relative energy spectrum. The points are the data, 
the thick solid line the result of a fit including an $s$-wave resonance
(thin solid line) and a $d$-wave resonance (dashed line) and in the right 
panel, an event-mixing "background" (dotted-dashed line). The parameters shown are 
those of the $s$-wave resonance Breit-Wigner lineshape. Note : a third resonance 
was tentatively introduced near 4 MeV (dotted line) but is not statistically significant. 
Its presence does not modify the fit in the region of interest.}
\label{fig:13Be}
\end{center}
\end{figure}

The relative energy between the $^{15}$B ($^{12}$Be) fragment and the neutron detected in coincidence was reconstructed event by event and is shown 
in figure \ref{fig:16B} (\ref{fig:13Be}). In the case of $^{16}$B ($^{15}$B+$n$ coincidences) a strong, narrow peak appears at about 100 keV above neutron 
threshold. For $^{13}$Be, one can distinguish two main features: a broad peak around 700 keV and a second structure at about 2 MeV.

\begin{figure}
\begin{center}
\mbox{\psfig{file=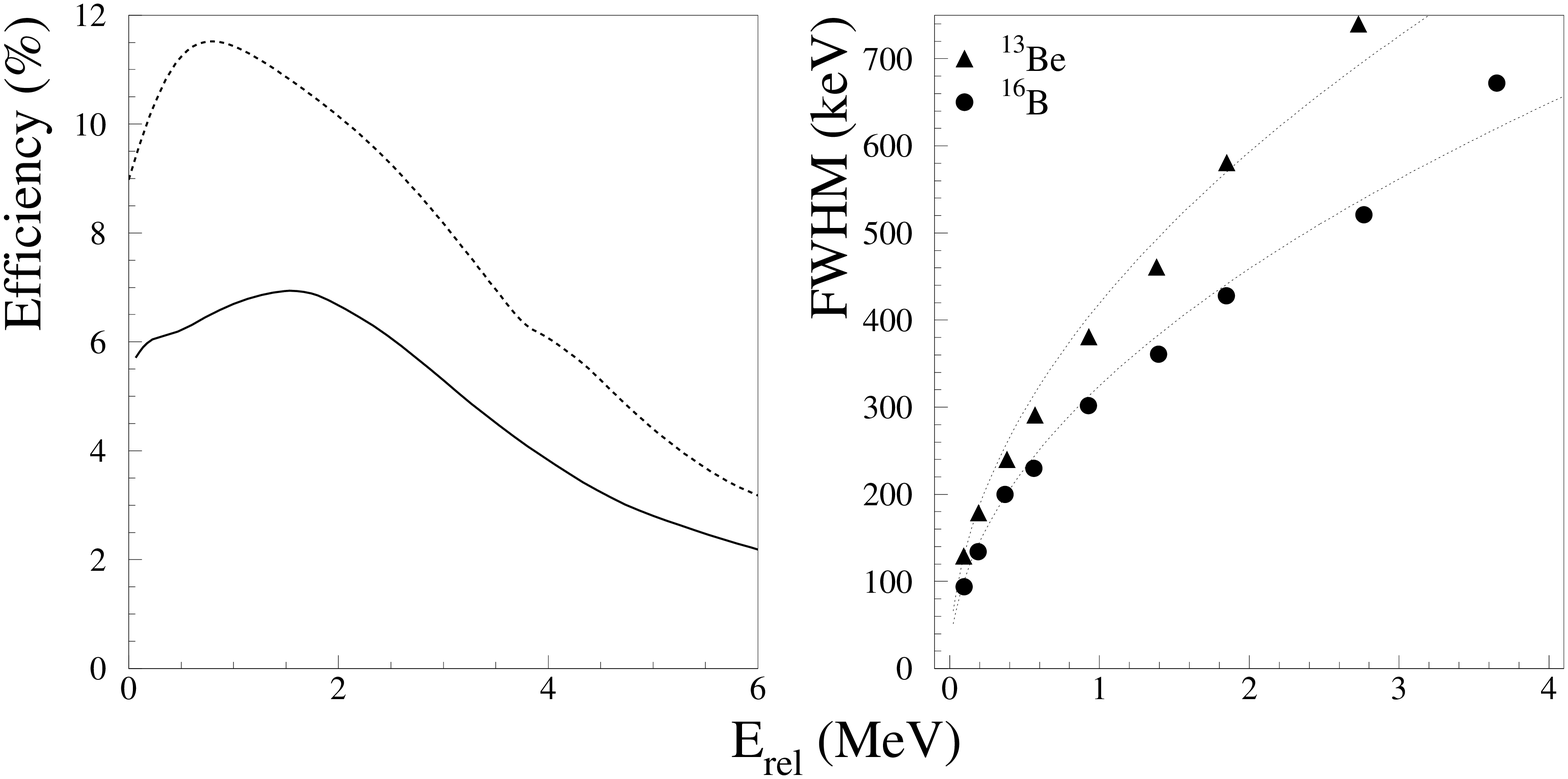,width=14cm}}
\caption{Setup efficiency and resolution simulated using GEANT. 
Left panel: detection efficiency as a function of relative energy 
between the fragment (solid line: $^{15}$B, dashed line: $^{12}$Be) and 
the neutron. Right panel: relative energy resolution as a function of relative energy. 
The curves are fits of the form FWHM $\propto \sqrt{E_{rel}}$.}
\label{fig:simul}
\end{center}
\end{figure}

To proceed further it is necessary to take into account the distorsions caused by the finite resolution and acceptance of the detectors. The response of the 
complete setup was simulated using the GEANT package \cite{Gea87}.
The resolution in relative energy (figure \ref{fig:simul}) which increases roughly 
as $\sqrt{E_{rel}}$ arises principally from the finite size of the neutron detectors.
The results also show that in both experiments the detection efficiency is 
a smooth function of the relative energy, ruling out the possibility that the peaks 
are instrumental artifacts. 

Energy distributions corresponding to uncorrelated
fragment-neutron pairs were generated by mixing neutrons and $^{15}$B ($^{12}$Be) fragments 
arising from different events \cite{Mar00}.
In both cases the lineshapes (figures \ref{fig:16B} and \ref{fig:13Be}) cannot describe the experimental spectra, which demonstrates that the structures
obtained are not due to a trivial phase space effect and correspond rather to fragment-neutron final state interaction.

For the interpretation of the data, we have employed the same formalism as in ref. \cite{Che01}. Starting from the sudden approximation, the relative
energy distribution is given by the overlap of the initial bound-state wave function, describing the relative motion between a neutron and the rest
of the projectile, and the final-unbound state wave function (neutron-fragment motion). Realistic wave functions were obtained by using Woods-Saxon
potentials adjusted to reproduce either the projectile neutron separation energy for the intial state or the resonance energy
(or the scattering length for virtual $s$-states) for the final state. The theoretical 
energy distributions were then folded with the simulated
response of the experimental setup.

As mentioned earlier, to first order the neutron configuration of the projectiles 
is preserved in proton knockout. The structures of $^{17}$C and of $^{14}$B are 
relatively well known to be mainly \mbox{$|^{16}$C$^*$(2$^+$)$>$} $\otimes$ $\nu d_{5/2}$ ($\ell$=2) and 
\mbox{$|^{13}$B(gs)$>$} $\otimes$ $\nu sd$ ($\ell$=0,2). Hence only $d$ states for $^{16}$B and $s$ and $d$ states for $^{13}$Be should be significantly populated in our experiments.

The $^{16}$B spectrum was fit with a lineshape corresponding to a $d$-wave resonance and the distribution obtained by event-mixing which
was the best description of the ``background'' lying below the narrow peak in the 
spectrum\footnote{ This background could be due to other ovelapping
$^{16}$B states or to uncorrelated $^{15}$B--neutron events (continuum).}. 
The resonance energy $E_r$ and the ratio between the two components were allowed
to vary freely. The result obtained was $E_r$=85$\pm$15 keV (above neutron threshold).
 At this energy the single-particle width is very small
($\sim$0.5 keV) and is dominated by the experimental resolution ($\sim$100 keV).

This resonance can most probably be identified as the ground state of $^{16}$B. In terms of 
the shell model the very small spectroscopic factor for neutron decay (0.08) of 
the predicted $0^-$ ground state would
be responsible for such a very narrow width \cite{Lec03}. This result should furnish a first 
constraint
on the $^{15}$B-$n$ interaction used in three-body calculations of the structure of $^{17}$B.

Turning to $^{13}$Be, the structure seen at around 2 MeV was identified as the $d$-wave resonance seen in previous experiments \cite{Ost92,Kor95,Bel98}.
The broad peak at lower energy requires more attention. As its large width is 
incompatible with a $d$-wave resonance at this energy, only an $s$-state
remains possible. Hence a fit was first attempted including a virtual $s$-state, 
a $d$-wave resonance and a ``background'' resulting from event-mixing. It was 
found, however, that such a prescription could not adequately describe the data \cite{Lec02}. 

In the simple model considered so far, no $s$-wave \textit{resonance} can exist owing to the absence of a centrifugal barrier. However $^{12}$Be, is well known
to be a deformed nucleus in which $N=8$ is no longer magic \cite{Iwa00}. It is likely, 
therefore,
that $^{12}$Be is not an inert core in $^{13}$Be, which might explain the appearance of
\textit{resonant} $s$-states.

A second fit was thus carried out with the $d$-wave resonance but replacing the virtual $s$ state by a resonance described by an $\ell=0$ Breit-Wigner lineshape. 
The results are shown in figure \ref{fig:13Be}. The agreement with the data is 
nearly perfect. Unlike the $^{16}$B case, the magnitude of the ``background''
could not be uniquely determined (the two-fold figure shows the two extreme cases) and 
the resonance parameters are consequently difficult to deduce exactly.
Nevertheless this experiment confirms the presence of a broad $s$-state at low energy 
(about 700 keV above threshold) in $^{13}$Be and, consequently, that 
the $s_{1/2}$-$d_{5/2}$ inversion in the N=9 isotones $^{14}$B and $^{15}$C persists in $^{13}$Be.
We note that such a low-lying $s$ state was also seen in a recent experiment in GSI
\cite{Sim03}, in contradiction with the earlier observation of a very low-lying 
virtual $s$-state \cite{Thn00}.
More theoretical work and experiments such 
as \mbox{$d$($^{12}$Be,$^{13}$Be)$p$} would help clarifying the structure of $^{13}$Be. 

\bigskip

   The author would like to thank the members of the E281a and E378 collaborations and
   the Group ``Noyaux Exotiques'' at LPC-Caen for their involvement in the work
   presented here.  The assistance of the technical staffs of LPC-Caen and GANIL in
   preparing and executing the experiments is gratefully acknowledged.

\end{document}